\documentclass[showpacs, pra,twocolumn,preprintnumbers ,amsmath, amssymb, superscriptaddress, aps]{revtex4-2}

\usepackage[utf8]{inputenc}
\DeclareUnicodeCharacter{202F}{\,}
\usepackage{color}
\usepackage{amsmath,amssymb}
\usepackage{pifont}
\usepackage{amssymb}  
\usepackage{bbold}
\usepackage{float}
\usepackage{subfloat}
\usepackage[squaren]{SIunits}

\usepackage[caption=false]{subfig}
\usepackage{tikz}
\usepackage{makecell}
\usepackage{subfig}
\usepackage{pifont}   
\usepackage{graphicx} 
\graphicspath{{Figures/}}
\usepackage{dcolumn}  
\usepackage{bm}       
\usepackage{multirow} 
\usepackage{placeins}
\usepackage[colorlinks]{hyperref}
\usepackage{mathtools}
\usepackage{appendix}

\begin{document}
	
	\title{Floquet-driven tunneling control in monolayer MoS$_2$}

	\date{\today}
	
	\author{Rachid El Aitouni}
	\affiliation{Laboratory of Theoretical Physics, Faculty of Sciences, Choua\"ib Doukkali University, PO Box 20, 24000 El Jadida, Morocco}
	
	\author{Aotmane En Naciri} 
	\affiliation{LCP-A2MC, Université de Lorraine, ICPM, 1 Bd Arago 57070 Metz, France}
	\author{Clarence Cortes}
	\affiliation{Vicerrector\'ia de Investigaci\'on y Postgrado, Universidad de La Serena, La Serena 1700000, Chile}  
	\author{David Laroze}
	\affiliation{Instituto de Alta Investigación, Universidad de Tarapacá, Casilla 7D, Arica, Chile}

\author{Ahmed Jellal}
\email{a.jellal@ucd.ac.ma}
\affiliation{Laboratory of Theoretical Physics, Faculty of Sciences, Choua\"ib Doukkali University, PO Box 20, 24000 El Jadida, Morocco}

	\begin{abstract}

		We study how fermions in molybdenum disulfide MoS$_2$ interact with a laser field and a static potential barrier, focusing on the transmission probability.
		Our aim is to understand and control photon-assisted quantum transport in this two-dimensional material under external driving.
		We use the Floquet approximation to describe the wave functions in the three regions of the system. By applying continuity conditions at the boundaries, we obtain a set of equations involving an infinite number of Floquet modes. We explicitly determine transmissions involving the central band $E$ and the first sidebands $E \pm \hbar\omega$. As for higher-order bands, we use the transfer matrix approach together with current density to compute the associated transmissions.
		Our results reveal that the transmission probability oscillates for both spin-up and spin-down electrons. The oscillations of spin-down electrons occur over nearly twice the period of spin-up electrons. Among all bands, the central one consistently shows the highest transmission. We also find that stronger laser fields and wider barriers both lead to reduced transmission.
	Moreover, laser irradiation enables controllable channeling and filtering of transmission bands by tuning the laser intensity and system parameters.
	This highlights the potential of laser-driven MoS$_2$ structures for highly sensitive electromagnetic sensors and advanced optoelectronic devices.

	\end{abstract}
	
	\maketitle

	
	
	\section{Introduction}	\label{Intro}

	The graphene discovery in 2004 was a significant milestone, which revealed a new class of two-dimensional (2D) material-based technologies \cite {Novos2004}. Since that time, great attention has been paid to graphene in view of its novel and fantastic characteristic behaviors. It is flexible and mechanically robust \cite{prop}. It is also highly transparent, losing only $2.3\%$ of the light shining on it \cite{absor}, which makes it attractive for applications such as flexible displays and transparent electronics. Yet, despite these exciting properties, graphene currently exists alongside important limitations that prevent its use in practical electronic devices. The fundamental problem, however, is that graphene lacks a natural band gap. In materials that have a band gap \cite{zero,zero1}, current can be turned on and off derivative to an originating potential, akin to a standard semiconductor. In graphene, electrons are not really massive, and they can continue to flow without heating for a long period of time under an external field \cite{masless}. This has made it difficult to utilize graphene for basic electronic components like transistors. To deal with this problem, several techniques were developed to open a band gap in graphene. In one type of behavior, electrons are localized by fixed (static) or time-dependent (oscillating) potential barriers \cite{oscil1,oscil2,timepot,timepot2,doubletemps}. These barriers are able to modify the energy surface for electrons, contributing to the formation of the energy separation between valence and conduction bands. This method is effective in a vacuum and also in ideal experiments, but it leads to a new problem, the Klein tunneling. Thus, because of this quantum effect, electrons in graphene no longer get reflected coming to a potential barrier (in particular with a head-on incidence) \cite{klienexp,klien1,klien2}. Therefore, electrons can still move freely even at an interface with a gap, making it difficult to control the electronic transport.

The successful exfoliation of graphene has sparked a widespread search for other 2D materials. This quest recently encouraged the experimental discovery of stable 2D materials, i.e., molybdenum disulfide ($\text{MoS}_2$) \cite{Radisavljevic2011} and tungsten diselenide ($\text{WSe}_2$) \cite{Wang2012}. These materials have, similar to graphene, a hexagonal crystal structure. However, they are much smaller than a nanometer thick, usually only a few atomic layers thick \cite{Chhowalla2013}. But $\text{MoS}_2$ in its single-layer structure—unlike graphene—has no band gap, and in the monolayer form $\text{MoS}_2$ has a direct band gap \cite{Mak2010}. This makes $\text{MoS}_2$ more suitable to the application in electronic and optoelectronic devices. Another important attribute of $\text{MoS}_2$ is its strong spin-orbit coupling splitting the valence bands at $K$ and $K'$ points of the Brillouin zone \cite{Zhu2011}. This results in the physics of coupled spin and valley. Together with the pseudospin, this renders $\text{MoS}_2$ a promising candidate for valleytronics, enabling the encoding and processing of information in the valley index of electrons \cite{Xu2014}. This class of properties makes $\text{MoS}_2$ an interesting material in novel devices beyond conventional charge-driven electronics.

	We investigate the tunneling of electrons through a barrier of width $D$ in a single layer of molybdenum disulfide MoS$_2$ subjected to a laser field with amplitude $A_0$ and frequency $\omega$. This study is motivated by our previous work on graphene, in particular \cite{Elaitouni2023A,ELAITOUNI2024,Elaitouni2023, Elaitouni2025,doublelaser}. We focus on how this field influences electron transport. We use the Floquet approximation to describe the interaction between the electrons and the laser field. This approximation allows us to express the wave functions in the three regions of the system in terms of time-periodic solutions. Applying the continuity conditions at the barrier boundaries yields a set of equations including an infinite number of Floquet modes. To make the problem manageable, we focus on the most relevant contributions: the central energy band ($E$) and the first sidebands ($E \pm \hbar\omega$). These bands play a dominant role in quantum transport under laser irradiation. We calculate the transmission probability for these cases using an analytical approach. For higher-order sidebands, for which an analytical treatment is difficult, we use a matrix formalism to efficiently and systematically compute the associated transmissions. This hybrid approach provides a more complete view of the transmission spectrum.

	We find that the maximal transmission is always provided by the central energy band for both spin-up and spin-down electrons. It presents an oscillating behavior based on system parameters such as the barrier's width and incoming particles' energy. This oscillation is due to quantum interference—when differently phased electron waves interfere with each other depending on their phase, which is affected by the laser field and the shape of the barrier. As the laser intensity increases, the transmission process becomes more complex. The interaction between the electrons and photons is also stronger with a stronger laser, which shifts the energy levels of the electrons, causing inelastic scattering. This scattering decreases total transmission due to making electron wave functions wider and resonance conditions shifted. Moreover, the laser may trap electrons in particular energy states due to the Stark effect \cite{Stark}. Being the effect for which quantum states become localized in a region induced by the action of an external field—recall that through it, electrons may have a probability and then be capable of going through the potential barrier. .
	
Our primary motivation is to elucidate how external time-periodic driving can be exploited to control quantum transport in 2D materials. In particular, we aim to demonstrate that laser irradiation provides an efficient and tunable mechanism to engineer photon-assisted transmission through electrostatic barriers in monolayer MoS$_2$. By analyzing the role of Floquet sidebands, spin-dependent oscillations, and barrier parameters, our calculations establish a direct link between microscopic quantum dynamics and experimentally accessible transport signatures. This perspective clarifies the relevance of the present study for laser-controlled filtering, band selectivity, and the design of optoelectronic and sensing devices based on driven van der Waals materials.

The present paper is organized as follows. In Sec. \ref{Theory}, we present our model and determine the wave functions corresponding to each region and each value. In Sec. \ref{TTMM}, we analyze the transmission probability in the central band as well as in the first lateral band. For the other modes, we employ the matrix formalism. 
We present and discuss our numerical analysis in Sec. \ref{NNRR}. Finally, we summarize and conclude our results.

	
	\section{Theoretical Model}	\label{Theory}
	
\begin{figure}[H]
		\centering
		\includegraphics[scale=0.5]{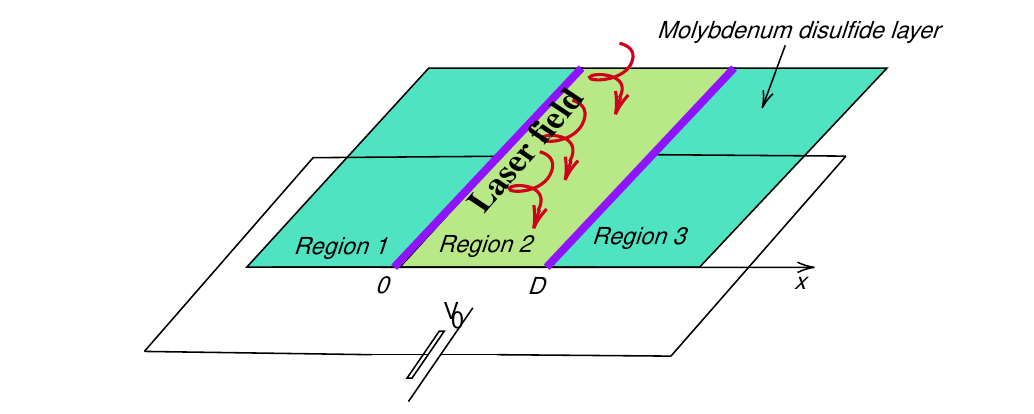}
		\caption{
	A schematic of a monolayer MoS$_2$  sheet is shown. A static potential barrier of height $V_0$ is applied to the central region of width $D$, which is exposed to a laser field with amplitude $A_0$ and frequency $\omega$.
		}\label{str}
\end{figure}

	We study a barrier structure applied to a monolayer molybdenum disulfide MoS$_2$ sheet composed of three regions $j=1,2,3$. The  static potential barrier $V_j(x)$ has a height $V_0$ and extends over a region of width $D$. A monochromatic laser field with linear polarization shines on this region. This setup changes the behavior of electrons in the material. The MoS$_2$ sheet is divided into three parts. The first part is the region before the barrier. The second part is the region where the barrier and the laser field are applied. The third part is the region after the barrier. This structure is shown in Fig. \eqref{str}. In this work, we look at how electrons move through the barrier under the influence of the laser. Our goal is to understand how the barrier and the light field affect the way electrons are transmitted. This model helps us explore how laser light can control electronic properties in 2D materials like MoS$_2$.

At low energy, the dynamics of the electrons in monolayer MoS$_2$ can be modeled with an effective Hamiltonian. It describes qualitatively the electronic structure of MoS$_2$ around the $K$ and $K'$ Dirac points of the Brillouin zone. It also contains important effects, like the direct band gap and strong spin-orbit coupling. The Hamiltonian is a good start in order to understand the behavior of electrons in the presence of external fields, i.e., electric, magnetic, and irradiation fields. The Hamiltonian of our model can be written as
\begin{widetext}
	\begin{equation}
	H_j =  v_F \left( \tau \sigma_x p_x + \sigma_y \left[p_y+\frac{e }{\hbar}A_j(t)\right] \right) + \frac{\Delta}{2}\sigma_z+ \frac{1-\sigma_z}{2}\lambda \tau s_z +V_j(x)
\end{equation}
\end{widetext}
where $p_x$ and $p_y$ are the components of the momentum vector, $\sigma_i$ ($i=x,y,z$) are the Pauli matrices, $\tau=1(-1)$ for the $K(K')$ valley, $s=1(-1)$ for spin-up (down), $v_F=at/\hbar$ is the Fermi velocity, $\Delta=1.887$ eV is the gap band, and $\lambda=0.082$ eV is the spin-orbit coupling \cite{Value}. The static potential barrier $V_j(x)$ of height $V_0$ and width $D$ is such that it is zero in regions $(j=1,3)$, while it is $V_0$ in region $j=2$, as indicated in Fig. \ref{str}. In the dipole approximation \cite{dipole}, the vector potential $A_j(t)$, generated by the laser field, is given by
\begin{align} 
	A_j(t)=\begin{cases}
		A_0 \cos(\omega t), &0<x<D\\
		0,& \text{otherwise}
	\end{cases}
\end{align} 
where $A_0$ denotes  the amplitude and $\omega$ is the frequency.
We can write the Hamiltonian in matrix form as
\begin{widetext}
	\begin{align}
	H_j =
	\begin{pmatrix}
		\frac{\Delta}{2}+V_j& v_F (\tau p_x - i \left[p_y+eA_0cos(\omega t)\right] \\
		v_F (\tau p_x + i \left[p_y+eA_0cos(\omega t)\right] & -\frac{\Delta}{2}+V_j +\lambda \tau s
	\end{pmatrix}.
\end{align}
\end{widetext}

In the presence of a laser field, the electron wave functions at the two valleys and the MoS$_2$ band structure undergo significant modifications, which are treated using the Floquet approximation \cite{approx}. Assuming a continuous-wave laser field, the spinors can be expressed as 
$\Psi^j(x,y,t)=\psi^j(x,y,t)e^{-iEt/\hbar}$
where $E$ is the Floquet quasi-energy, $\psi^j(x,y,t)$ is a time periodic function such that   $\psi^j(x,y,t+T)=\psi^j(x,y,t)$ with $T$ the time period of the laser field. By the translational invariance of the Hamiltonian in the $y$-direction,
the spinors can also be split as $\Psi^j(x,y,t)=\binom{\varphi_{c}^j(x)}{\varphi_{v}^j(x)} e^{ik_yy} f(t)$. The eigenvalue equation $H_j \Psi^j = E \Psi^j$ yields the two coupled equations  
\begin{widetext}
	\begin{align}
i \hbar v_F\left[\tau \frac{\partial}{\partial x} - \left(k_y + \frac{eA_0}{\hbar} \cos (\omega t)\right) \right]\varphi_v^j(x) f(t) &= \left[i \hbar  \frac{\partial f(t)}{\partial t} + f(t) \left( E -V_j- \frac{\Delta}{2} \right)\right] \varphi_c^j(x) \label{CCC}\\
i \hbar v_F\left[ \tau \frac{\partial}{\partial x} + \left(k_y + \frac{eA_0}{\hbar} \cos (\omega t)\right) \right]\varphi_c^j(x) f(t) &= \left[i \hbar  \frac{\partial f(t)}{\partial t} + f(t) \left( E- V_j + \frac{\Delta}{2} -\lambda \tau s \right)\right] \varphi_v^j(x)\label{VVV}.
\end{align}
\end{widetext}
As an initial step, we consider that the wave functions $\varphi_c^j(x)$ and $\varphi_v^j(x)$ satisfy a set of coupled differential equations in the barrier region, assuming no laser field is present. Within this approximation, \eqref{CCC} and \eqref{VVV} can then be used to derive
\begin{align}
&-\frac{ev_FA_0}{\hbar} \cos(\omega t) \varphi_v^j(x) f(t) =\varphi_c^j(x) \frac{\partial f}{\partial t} \\
&\frac{ev_FA_0}{\hbar} \cos(\omega t) \varphi_c^j(x) f(t) =\varphi_v^j(x) \frac{\partial f}{\partial t} 
\end{align}
which, in turn, yield the  second-order differential equation 
\begin{equation}
	 \frac{d^2 f}{d t^2} + \omega \tan(\omega t) \frac{d f}{d t} + \left(\frac{ev_FA_0}{\hbar}\right)^2 \cos^2(\omega t) f(t) = 0 
	 \end{equation}
 whose solution takes the form
\begin{align}
	f(t)\sim e^{-i \alpha sin(\omega t)}
\end{align}
where the dimensionless parameter $\alpha = \frac{ev_F A_0}{\hbar \omega}$ characterizes the strength of the electron–laser interaction.
To proceed, we use the Anger expansion formula to introduce Bessel functions of the first kind $J_m(\alpha)$
\begin{align}
	f(t)=\sum_{m=-\infty}^{+\infty}J_m(\alpha)e^{-i m \omega t}.
\end{align}
This decomposition is central to the Floquet analysis and reveals how the original time dependence can be treated as a sum over discrete harmonics, each weighted by $J_m(\alpha)$. It also reflects the possibility of photon-assisted processes, where electrons absorb or emit $m$ photons with energy $\hbar \omega$.
Finally, the spinors 
can be written as
\begin{equation}
	\Psi^j(x,y,t)=\binom{\varphi_{c}^j(x)}{\varphi_{v}^j(x)} e^{ik_yy}
	\sum_{m=-\infty}^{\infty}J_m(\alpha)e^{-i(E+m\omega \hbar) t/\hbar}.
\end{equation}
Injecting this equation into  (\ref{CCC}-\ref{VVV}) and use 
the relation
$\frac{\partial }{\partial t}e^{-i \alpha\sin(\omega t)}=-i\frac{ev_FA_0}{\hbar}\cos(\omega t) e^{-i \alpha \sin(\omega t)}$, to show that
the components $\varphi_{c}^j(x)$ and $\varphi_{v}^j(x)$ satisfy the equations
\begin{widetext}
		\begin{align}
&			\left[V_0 + \frac{\Delta}{2} - (E+\l\hbar\omega)\right]\varphi_c^j(x) - i\hbar v_F\left[ \tau \partial_x+ \left(k_y+\frac{l \omega}{v_F}\right)\right]\varphi_v^j(x)=0\\
	&		i\hbar v_F	\left[ -i\tau \partial_x +  \left(k_y+\frac{l \omega}{v_F}\right) \right]\varphi_c^j(x)+\left[V_0 - \frac{\Delta}{2} + \lambda \tau s - (E+l\hbar\omega)\right]\varphi_v^j(x)=0.
	\end{align}
\end{widetext}
	These  can be simplified and combined into a single second-order differential equation. Solving this equation in region 2 allows us to determine the eigenspinor
	\begin{widetext}
		\begin{align}
		\Psi^2(x,y,t)=\sum_{m,l-\infty}^{\infty}\left[a^l_2\begin{pmatrix}
			1\\ e^{i\varphi_l}
		\end{pmatrix}e^{i\tau q^l_xx}+b^l_2\begin{pmatrix}
		1\\ -e^{-i\varphi_l}
		\end{pmatrix}e^{-i\tau q^l_xx}\right]e^{-ik_yy}e^{-i(E+l\hbar\omega)t/\hbar}J_{m-l}(\alpha)
	\end{align}
	\end{widetext}
where  the wave vector component  in the direction of motion is given by
\begin{widetext}
	\begin{equation}
\left(q^l_x\right)^2 =\frac{1}{\hbar^2 v_F^2}\left(E+l\hbar\omega-V_0-\frac{\Delta}{2} \right)
\left(E+l\hbar\omega+\frac{\Delta}{2}-V_0-\lambda\tau s\right)-\left(k_y+\frac{l \omega}{v_F}\right)^2
\end{equation}
\end{widetext}
and the angle is  $\varphi_l=\tan^{-1}\frac{k_y+l\omega/v_F}{q^l_x}$. 
As for regions 1 and 3, the eigenspinors can be obtained as follows
\begin{widetext}
		\begin{align}
	&\Psi^1(x,y,t)=\sum_{m,l-\infty}^{\infty}\left[\delta_{l,0}\begin{pmatrix}
		1\\ e^{i\theta_l}
	\end{pmatrix}e^{i\tau k^l_xx}+r_l\begin{pmatrix}
		1\\ -e^{-i\theta_l}
	\end{pmatrix}e^{-i\tau k^l_xx}\right]e^{-ik_yy}e^{-i(E+l\hbar\omega)t/\hbar}\delta_{m,l}\\
	&
	\Psi^3(x,y,t)=\sum_{m,l-\infty}^{\infty}t_l\begin{pmatrix}
		1\\ e^{i\theta_l}
	\end{pmatrix}e^{i\tau k^l_xx}e^{-ik_yy}e^{-i(E+l\hbar\omega)t/\hbar}\delta_{m,l}
\end{align}
\end{widetext}
such that  the wave vector component reads as
\begin{widetext}
\begin{align}
	\left(k^l_x\right)^2 =\frac{1}{\hbar^2 v_F^2}\left(E+l\hbar\omega-\frac{\Delta}{2} \right)
	\left(E+l\hbar\omega+\frac{\Delta}{2}-\lambda\tau s\right)-k_y^2 
\end{align}
\end{widetext}
and  the angle	 is $\theta_l=\tan^{-1}\frac{k_y}{k^l_x}$. 
In the following, we will demonstrate how the above results can be employed to analyze the tunneling behavior through the present system, particularly by examining how the transmission probabilities vary with relevant physical parameters such as energy, barrier width, laser intensity, and incident angle.

\section{Transmission modes}\label{TTMM}

In order to analyze how electrons move through the MoS$_2$ sheet, we begin by carefully applying boundary conditions to the edges of the potential barrier. These conditions ensure that the wave functions describing the electrons connect smoothly from one region to another, with no abrupt changes. Along with this, we use the concept of current density to track the flow of electrons through the system. Combining these two approaches—the boundary conditions and the current density—allows us to accurately determine the transmission modes. These modes give the transmission and reflection probabilities for the electron to pass through or be reflected back by the barrier. Such an approach is necessary in order to unravel the transport behaviors of MoS$_2$ in different scenarios and to estimate its performance as well as potential applications in practical electronics and optoelectronics. 
Then, by imposing continuity at the interfaces $(x = 0, D)$, we write
\begin{align}
	&	\Psi^1(0,y,t) = \Psi^2(0,y,t) \\
	& \Psi^2(D,y,t) = \Psi^3(D,y,t)
\end{align}
and taking into account the orthogonalization term $e^{i v_F m \varpi t}$, we  obtain 
\begin{align}
	&\delta_{m,0}+r_m=\sum_{l=-\infty}^{\infty}\left(a_l+b_l\right)J_{m-l}\label{2121}\\
	&\delta_{m,0}e^{i\theta_m} -r_me^{-i\theta_m}=\sum_{l=-\infty}^{\infty}\left(a_l
	e^{i\varphi_l}
	-b_le^{-i\varphi_l}\right)J_{m-l}\label{2222}\\
	&t_me^{i\tau k_mD}=\sum_{l=-\infty}^{\infty}\left(a_le^{i\tau q_lD}+b_le^{-i\tau q_lD}\right)J_{m-l}\label{2323}\\
	&t_m e^{i\theta_m}
	e^{i\tau k_mD}=\sum_{l=-\infty}^{\infty}\left(a_le^{i(\tau q_lD+\varphi_l) }-b_le^{-i(\tau q_lD+\varphi_l)}\right)J_{m-l}.\label{2424}
\end{align}
Here, we have adopted the notation $J_{m-l} = J_{m-l}(\alpha)$. To proceed, we first restrict our analysis to the first three modes and explicitly derive the corresponding transmission probabilities. Specifically, for $m = 0$ (central band), the above equations reduce to
\begin{align}
	&1+r_0=\left(a_{0}+b_{0}\right)J_{0}\\
	&e^{i\theta_0}-r_0e^{-i\theta_0}=\left(a_{0}e^{i\varphi_0}-b_{0}e^{-i\varphi_0}\right)J_{0}\\
	&t_0e^{i\tau k_0D}=\left(a_{0}e^{i\tau q_0D}+b_{0}e^{-i\tau q_0D}\right)J_{0}\\
	&t_0 e^{i(\tau k_0D+ \theta_0)}=\left(a_{0} e^{i(\tau q_0D+ \varphi_0)}-b_{0}e^{-i(\tau q_0D+\varphi_0)}\right)J_{0}.
\end{align}
Since the zero mode ($m = 0$) is coupled to the other modes ($m = \pm 1$), we need to determine all the coefficients associated with $m = 0$. 
After carrying out some algebraic manipulations, we arrive at the final results
\begin{widetext}
	\begin{align}
    	&t_0=\frac{
		e^{-i D \tau k_x^0} \cos\theta_0 \cos\varphi_0	}{ \cos(D \tau q_x^0) \cos\theta_0 \cos\varphi_0- 
		i \sin(D \tau q_x^0) (1 -  \sin\theta_0 \sin\varphi_0)}\label{21}\\
        &	r_0=\frac{i e^{-i (D \tau q_x^0 + \varphi_0)} (-1 + e^{2 i D \tau q_x^0}) (e^{i \varphi_0}- e^{i \theta_0}) (1+ e^{i (\Lambda^+_{0,0})} )}{
		4 \left[ i \cos(D \tau q_x^0) \cos\theta_0 \cos\varphi_0 + \sin(D \tau q_x^0) \left( 1- \sin\theta_0 \sin\varphi_0\right) \right]}\label{22}\\
        &a_0=\frac{i e^{-i (D \tau q_x^0 + \Lambda^+_{0,0})} (1 + e^{2 i \theta_0})(1 + e^{i (\Lambda^+_{0,0})})}{4 J_0 \left[i \cos(D \tau q_x^0) \cos\theta_0 \cos\varphi_0 +\sin(D \tau q_x^0) \left(1 -\sin\theta_0 \sin\varphi_0\right)\right]}\label{23}\\
        &b_0= \frac{i e^{i (D \tau q_x^0 - \theta_0)} (1 + e^{2 i \theta_0})(e^{i \varphi_0}- e^{i \theta_0})}{4 J_0 \left[i \cos(D \tau q_x^0) \cos\theta_0 \cos\varphi_0 + \sin(D \tau q_x^0) \left(1- \sin\theta_0 \sin\varphi_0 \right)\right]}\label{24}
\end{align}
\end{widetext}
where $\Lambda^+_{0,0}= \theta_0 + \varphi_0$. As a result, from 
\eqref{21}, we derive the transmission probability $T_0= |t|^2$ for the zero mode
\begin{widetext}
	\begin{align}
	T_0=
\frac {\cos^2\theta_0 \cos^2 \varphi_0}{\cos^2\theta_0  \cos^2\varphi_0 \cos^2 (D q_0 \tau)+\sin^2 (D q_0 \tau)\left(1- \sin\theta _0 \sin\varphi _0\right)^2}
\end{align}
\end{widetext}
which satisfies the normalization condition $T_0+ R_0=1$.

For $m=\pm1$, and with the approximation that the $m>\alpha$ terms can be dropped due to being much smaller, the equations (\ref{2121}-\ref{2424}) simplify accordingly. This enables us to consider the coupling terms between the central mode ($m = 0$) and its neighboring modes only. The corresponding system can be cast in the form
%
\begin{widetext}
	\begin{align}
		&r_{\pm 1}=\left(a_{0}+b_{0}\right)J_{{\pm 1}}+ \left(a_{{\pm 1}}+b_{{\pm 1}}\right)J_{0}\\
	&e^{i\theta_{\pm 1}}-r_{\pm 1}e^{-i\theta_{\pm 1}}=\left(a_{0}e^{i\varphi_0}-b_{0}e^{-i\varphi_0}\right)J_{{\pm 1}}+ \left(a_{{\pm 1}}e^{i\varphi_{\pm 1}}-b_{{\pm 1}}e^{-i\varphi_{\pm 1}}\right)J_{0}\\
	&t_{\pm 1}e^{i\tau k_{\pm 1}D}=\left(a_{0}e^{i\tau q_x^0D}+b_{0}e^{-i\tau q_x^0D}\right)J_{{\pm 1}}+ \left(a_{{\pm 1}}e^{i\tau q_{\pm 1}D}+b_{{\pm 1}}e^{-i\tau q_{\pm 1}D}\right)J_{0}\\
	&t_{\pm 1}e^{i(\tau k_{\pm 1}D+ \theta_{\pm 1})}=\left(a_{0} e^{i(\tau q_x^0D+ \varphi_0)}-b_{0}e^{-i(\tau q_x^0D+\varphi_0)}\right)J_{{\pm 1}}+ \left(a_{{\pm 1}} e^{i(\tau q_{\pm 1}D+ \varphi_{\pm 1})}-b_{{\pm 1}}e^{-i(\tau q_{\pm 1}D+\varphi_{\pm 1})}\right)J_{0}.
\end{align}
\end{widetext}
After a lengthy algebraic calculation, we obtain the corresponding transmission coefficients for the sidebands ($m = \pm 1$). These expressions reflect how the electron wavefunction interacts with the time-periodic potential and how the energy is redistributed among the modes. The resulting transmission coefficients are given by
\begin{widetext}
	\begin{align}
	t_{m}=&\frac{
		e^{-\frac{i}{2}(-\Theta^+_{0,m} + 2 d k_m \tau)} 
		J_m
		\cos\theta_0} {J_0 \left( 
		1 + e^{2 i d q_0 \tau} (-1 + \cos\Lambda^-_{0,0}) + \cos\Lambda^+_{0,0}\Big)\Big(e^{2 i d q_m \tau} \sin\Lambda^-_{m,m} + \sin\Lambda^+_{m,m}\right)}\times\notag\\		
& 
		\left[ \cos\varphi_m
		\left( 
		-2 e^{i d q_m \tau} (-1 + e^{2 i d q_0 \tau}) \sin\frac{\Theta^-_{0,m}}{2}
		+ (e^{i d q_0 \tau} - 2 e^{i2 d  Q^+_{0,m} \tau}) \sin\Phi^+_{0,m, -}
	 + (e^{i d q_0 \tau} - 2 e^{i d q_m \tau}) \sin\Phi^+_{0,m, +}
		\right)\right. \nonumber\\
		& \left.
		+ 2 e^{i d q_0 \tau} \cos\varphi_0
		\left(
		-\sin\frac{\Theta^-_{0,m}}{2}
		+ 2 e^{2 i d q_m \tau} \cos\frac{\Lambda^-_{m,m}}{2} \sin\frac{\Lambda^-_{0,m}}{2}
		+ \cos\frac{\Theta^+_{0,m}}{2} \sin\varphi_m
		\right)
		\right].
\end{align}
\end{widetext}
These coefficients allow us to determine the corresponding transmission probabilities using $T_m = |t_m|^2$, expressed as
\begin{widetext}
	{ \begin{align}
   & T_m=J_m^2 \cos^2\theta_0\Bigg[-4 e^{-i d Q^+_{0,m} \tau} \cos\varphi_m\Big([\cos(d q_0 \tau) - \cos(d q_m \tau)] \cos\varphi_0 \sin\frac{\Theta^+_{0,m}}{2}
            - i \sin(d q_0 \tau)
                \sin\frac{\Theta^-_{0,m}}{2}\nonumber\\
        &\times\Bigg\{-4 e^{i d Q^+_{0,m} \tau} \cos\varphi_m
        \Big([\cos(d q_0 \tau) - \cos(d q_m \tau)] \cos\varphi_0 \sin\frac{\Theta^+_{0,m}}{2}
            + i \sin(d q_0 \tau)\Big(\sin\frac{\Theta^-_{0,m}}{2}
                - \cos\frac{\Theta^+_{0,m}}{2} \sin\varphi_0
            \Big)\Big)\nonumber\\
            &+ 2 e^{i d q_0 \tau} \left(-1 + e^{2 i d q_n \tau}\right) \cos\varphi_0
        \Big(\sin\frac{\Theta^-_{0,m}}{2}-\cos\frac{\Theta^+_{0,m}}{2} \sin\varphi_m \Big)
    \Bigg\}\Bigg] \times \Bigg[J_0^2\Big(1 + e^{-2 i d q_0 \tau} (-1 + \cos\Lambda^-_{0,0}) + \cos\Lambda^+_{0,0}\Big)\nonumber\\
    &\times\Big(1 + e^{2 i d q_0 \tau} (-1 + \cos\Lambda^-_{0,0}) + \cos\Lambda^+_{0,0}\Big)\Big( e^{-2 i d q_m \tau} \sin\Lambda^-_{m,m} + \sin\Lambda^+_{m,m}\Big) \Big(e^{2 i d q_m \tau} \sin\Lambda^-_{m,m} + \sin\Lambda^+_{m,m}\Big)\nonumber\\
    &- \cos\frac{\Theta^+_{0,m}}{2} \sin\varphi_0
            \Big)\Big)
            + 2 e^{-i d q_0 \tau} \left(-1 + e^{-2 i d q_m \tau}\right) \cos\varphi_0
        \left(\sin\frac{\Theta^-_{0,m}}{2}-\cos\frac{\Theta^+_{0,m}}{2} \sin\varphi_m
        \right)\Big]^{-1}.
\end{align}}
\end{widetext}
These transmissions fulfill the normalization condition $T_m + R_m = 1$.
Here, we have introduced the following shorthand notations:
 $\Phi^+_{0,m, \pm}= \frac{\Theta^+_{0,m}\pm 2\varphi_0}{2}$, 
   $ \Lambda^\pm_{i,j}=\theta_i \pm \varphi_j$, 
     $\Theta^\pm_{i,j}=\theta_i \pm \theta_j$, and 
    $ Q^\pm_{i,j}=q_i\pm q_j$.


For higher-order modes ($m > 1$), we turn to the transfer matrix method combined with current density analysis to compute the transmission probabilities. First note that the system of equations given by  (\ref{2121}–\ref{2424}) contains an infinite number of coupled equations involving an infinite set of unknowns indexed by $m$ and $l$, each ranging from $-\infty$ to $+\infty$. Solving such an infinite system analytically is not feasible. To be able to handle the problem, we make a standard and bona fide approximation. We cut off the series and keep only a finite number of Floquet modes in the interval $-N$ to $N$, with the cutoff being determined by the coupling strength $\alpha$ \cite{M1999}. This selection is due to the behaviour of $J_m(\alpha)$ describing photon-assisted transitions in the Floquet formalism. Because $J_m(\alpha)$ decays quickly for $|m| > \alpha$, the higher order terms become unimportant. Thus this truncation serves as a practical realization of the relevant physics while making the system tractable. By applying this approximation and implementing the transfer matrix approach, we can compute the transmission modes for $m > 1$ and gain a more complete understanding of the transport behavior in laser-irradiated MoS$_2$ systems. Let us  rearrange the set  (\ref{2121}–\ref{2424}) as
\begin{align}\label{Tmat}
	\binom{\delta_{m,0}}{r_m}
	=\mathbb{M}\begin{pmatrix}
		t_m\\
		\mathbb{0}_m
	\end{pmatrix}
\end{align}
where $\mathbb{M}$  being the transfer matrix connecting  regions 1 and 3
\begin{widetext}
	\begin{align}
	\mathbb{M}= 	\begin{pmatrix}
		\mathbb{M}_{11}&	\mathbb{M}_{12}\\
		\mathbb{M}_{21}&	\mathbb{M}_{22}
	\end{pmatrix}=\begin{pmatrix}
	\mathbb{I}&\mathbb{I}\\
	\mathbb{N}^+&\mathbb{N}^-
	\end{pmatrix}^{-1}\cdot \begin{pmatrix}
	\mathbb{J}&\mathbb{J}\\
	\mathbb{G}^+&\mathbb{G}^-
	\end{pmatrix}.\begin{pmatrix}
	\mathbb{E}&\mathbb{E}\\
	\mathbb{D}^+&\mathbb{D}^-
	\end{pmatrix}^{-1}\cdot \begin{pmatrix}
	\mathbb{S}&\mathbb{S}\\
	\mathbb{C}^+&\mathbb{C}^-
	\end{pmatrix}\label{37}
\end{align}
\end{widetext}
	and different involved matrices have the  elements 
	\begin{align}
		&(\mathbb{N}^\pm)_{m,l}=\pm(\Gamma_m)^{\pm 1}\delta_{m,l}\\
        &(\mathbb{J}^\pm)_{m,l}=\pm J_{m-l}(\alpha)\\
        &(\mathbb{G}^\pm)_{m,l}=\pm \gamma^{\pm 1}_l J_{m-l}(\alpha)\\
		&(\mathbb{E}^\pm)_{m,l}=e^{\pm i k_x^l D}\\
        &(\mathbb{D}^\pm)_{m,l}=\pm \Gamma^{\pm 1}_l e^{\pm i k_x^l D}\\
        &(\mathbb{S}^\pm)_{m,l}=\pm J_{m-l}(\alpha)e^{\pm iq_x^lD}\\
		&(\mathbb{C}^\pm)_{m,l}=\pm J_{m-l}(\alpha)\gamma^{\pm1}_le^{\pm iq_x^lD}.
	\end{align}
	Thus, from  \eqref{Tmat} and \eqref{37}, we derive the  transmission coefficients for all modes as
		\begin{equation}
		t_m=\mathbb{M}_{1,1}^{-1}\cdot\delta_{m,0}
	\end{equation}
	where $m\in [-N,N]$.
	We will use the continuity equation in the present analysis to find the current densities of the incident and reflected waves and transmitted wave. This is an important step in enabling a simulation of electron behavior in the current through the barrier region. By ensuring current conservation at the boundaries, we can accurately determine the amount of current transmitted through or reflected by the potential and laser-irradiated regions of the MoS$_2$ sheet. This step completes the framework needed to calculate the transmission probabilities for each Floquet mode. We get the densities
	\begin{align}
	&	J_{i}^0=v_F(\gamma_0+\gamma^*_0)\\
	&	J_{t}^l=v_Ft^*_lt_l(\gamma_l+\gamma^*_l)\\
	&	J_{r}^l=v_Fr^*_lr_l(\gamma_l+\gamma^*_l).
	\end{align}
	By comparing the transmitted current to the incident one, we can evaluate the transmission probability for each mode $l$ as
	\begin{align}
		T_l=\frac{|J_{t}^l|}{|J_{i}^0|}=|t_l|^2.	
	\end{align}
The total transmission is then obtained by summing over all propagating modes
	\begin{equation}
		T=\sum_{l=-N}^{N}T_l.
	\end{equation}	
	These results offer valuable insight into the electron transport behavior of the system across different physical conditions. They show how barrier width, laser intensity, and incident energy determine whether electrons are transmitted or reflected. This information is crucial for the design of MoS$_2$-based devices in which regulating electron transport is paramount for tunable electronic or optoelectronic applications.  
	Next, we focus on the $K$ valley transmission to keep the presentation concise while highlighting the essential physics.

\section{NUMERICAL RESULT}\label{NNRR}

\begin{figure}[ht]
	\centering
	\subfloat[ \( \phi = 0^\circ \)]{\includegraphics[scale=0.55]{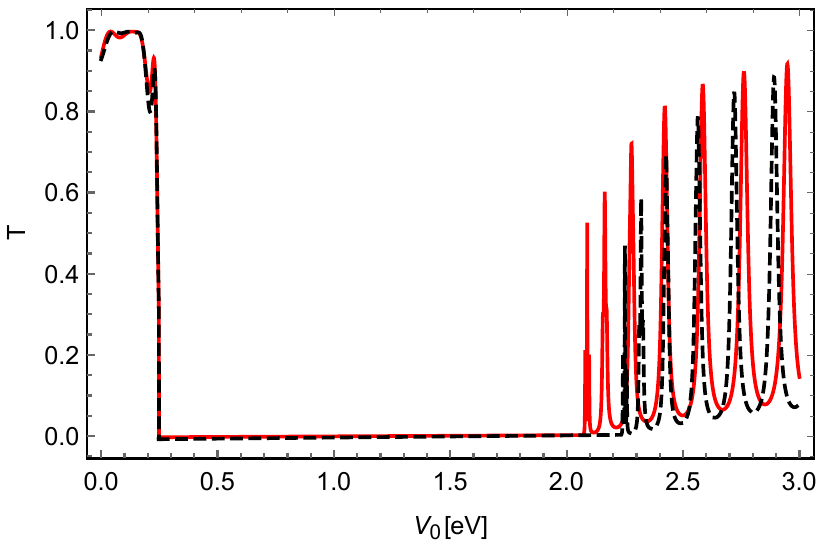}\label{fig4a}}
	\\
	\subfloat[ \( \phi = 30^\circ \)]{\includegraphics[scale=0.55]{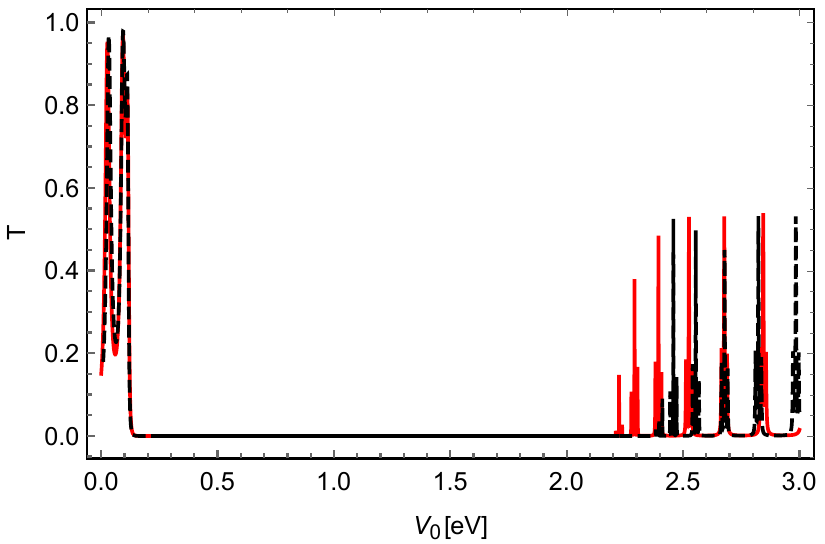}\label{fig4b}}
	\\
	\subfloat[ \( \phi = 45^\circ \)]{\includegraphics[scale=0.55]{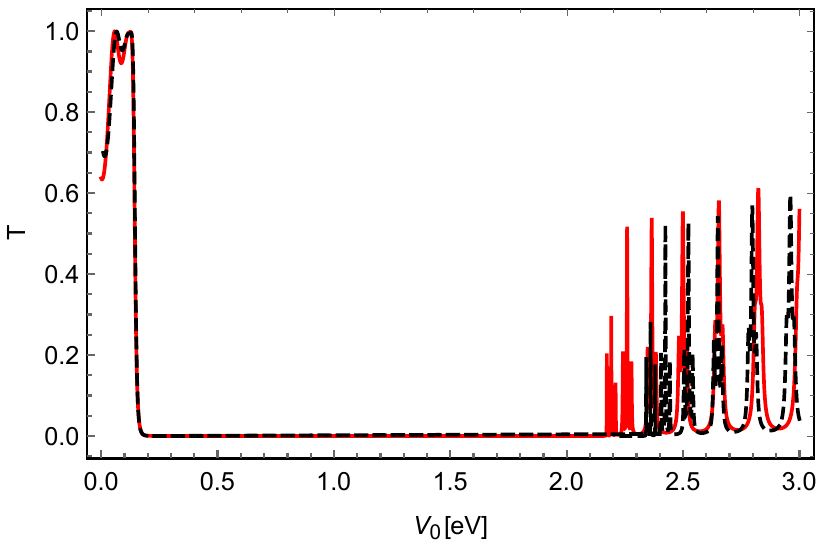}\label{fig4c}}
		\caption{Transmission spin-up  ($s_z=1$: red solid line) and spin-down ($s_z=-1$: black dashed line) for the valley $K$ as a function of the barrier height \( V_0 \) for three different incident angles $(\phi = 0^\circ, 30^\circ, 45^\circ)$, laser frequency 
				\( \omega = 15 \times 10^{12} \) Hz, parameter \( \alpha = \frac{ev_F A_0}{\hbar \omega}= 1 \),  incident energy \( E = 1.2 \) eV, and barrier width \( D = 5 \) nm.}
	\label{fig4} 
\end{figure}

Fig. \ref{fig4} shows the total transmissions of spin-up (red solid line) and spin-down (dashed black line) as a function of the applied potential $V_0$ for three incident angles. For normal incidence (Fig. \ref{fig4a}), these transmissions are almost the same for $V_0 < E - \frac{\Delta}{2}$, by virtue of the weak spin coupling in the conduction band \cite{CHENG2015}. In the region where $ E - \frac{\Delta}{2} < V_0 < E + \frac{\Delta}{2} - \lambda \tau s_z$, the transmission probability is zero because the longitudinal wave vector $q_x$ becomes imaginary, leading to the formation of a transmission gap and completely blocking the passage of fermions. For $V_0 > E + \frac{\Delta}{2} - \lambda \tau s_z$, a small shift is observed between the transmission of spin-up and spin-down due to the strong spin-orbit coupling in the valence band. As we have seen in the case without a laser \cite{CHENG2015}. In this latter case, the barrier becomes perfectly transparent. If laser irradiation is applied, the transparency of the barrier disappears, and the transmission amplitude becomes dependent on the angle of incidence. For non-normal incidences (Figs.~\ref{fig4b} and \ref{fig4c}), the transmission decreases compared to normal incidence. The incident angle is not affected by the coupling. 
In other words, the spin-orbit interaction depends on how high the barrier is—or how much energy the incoming electrons have. Transmission only takes place when the barrier height $V_0$ matches certain conditions.

\begin{figure}[ht]
	\centering
	\subfloat[$E=1.2$]{\includegraphics[scale=0.55]{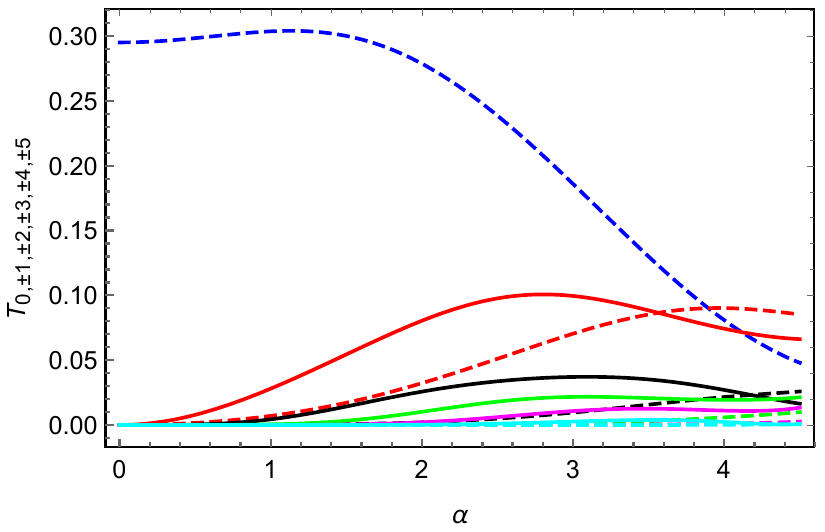}\label{fig5a}\includegraphics[width=0.7cm,height=4.9cm]{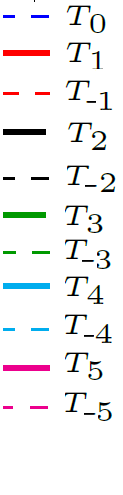}}\\
	\subfloat[$E=1.4$]{\includegraphics[scale=0.55]{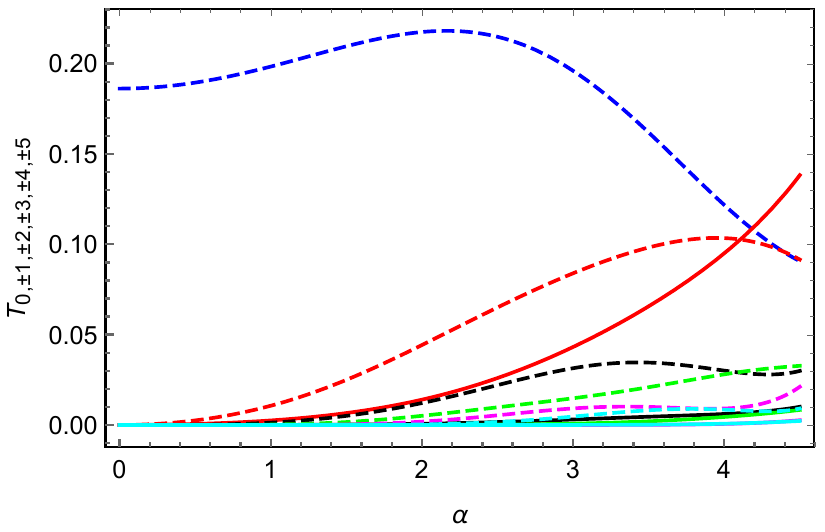}\label{fig5b}\includegraphics[width=0.7cm,height=4.9cm]{leg.png}}\\
	\subfloat[$E=1.6$]{\includegraphics[scale=0.55]{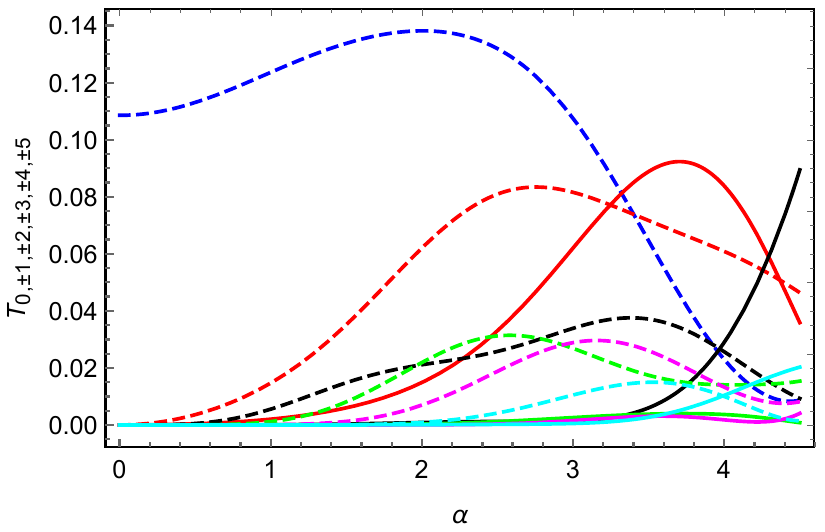}\label{fig5c}\includegraphics[width=0.7cm,height=4.9cm]{leg.png}}
	\caption{Transmissions spin-up ($s_z=1$) of the central band $T_0$ and the first five sidebands ($l=\pm1, \cdots,\pm 5$) as a function of the parameter $\alpha = \frac{ev_F A_0}{\hbar \omega}$ for three incident energies ($E=1.2$ eV, 1.4 eV, 1.6 eV),
		\(\omega = 15 \times 10^{12}\) Hz, $\phi=0$, $D=10$ nm, and $V=2.6$ eV} 
		\label{fig5}  
\end{figure}

Fig.~\ref{fig5} displays transmission without photon exchange  $T_0$ and transmissions with photon exchange $T_l$ ($l=\pm 1, \cdots, \pm 5$) as a function of the laser parameter $\alpha$. For $E=1.2$ eV, Fig.~\ref{fig5a} shows that at low irradiation, transmission is entirely governed by the central band, meaning no photon exchange occurs between the barrier and the fermions, as we have seen in the graphene case \cite{oscil1,oscil2,timepot}.
Transmissions involving photon exchange appear when the laser parameter $\alpha$ is increased. First, a single photon is exchanged to initiate transmission. But as $\alpha$ rises, more transmission modes emerge, involving the exchange of two photons, followed by three, four, and so forth \cite{Elaitouni2023A, ELAITOUNI2024, Elaitouni2023}.  
{As illustrated in Fig.~\ref{fig5b}, for an incident energy of $E = 1.4$ eV, transmission without photon exchange remains dominant at low values of $\alpha$, but gradually decreases as $\alpha$ increases. In parallel, transmission involving photon exchange becomes increasingly significant, particularly the emission of two or more photons, which is statistically more probable than absorption. Fig.~\ref{fig5c}, corresponding to an energy of $E = 1.6$ eV, confirms the previous trends. However, despite the increase in incident energy, the overall transmission amplitude decreases. These observations lead to the conclusion that increasing the laser irradiation parameter $\alpha$ promotes the emergence of transmission modes via photon exchange between the barrier and the fermions—a phenomenon already observed in the case of graphene. Photon interaction contributes to raising the energy of the fermions, thereby providing them with sufficient energy to cross the barrier.}

\begin{figure}[ht]
	\centering
	\subfloat[$\alpha=0$, n–p–n junction]{\includegraphics[scale=0.55]{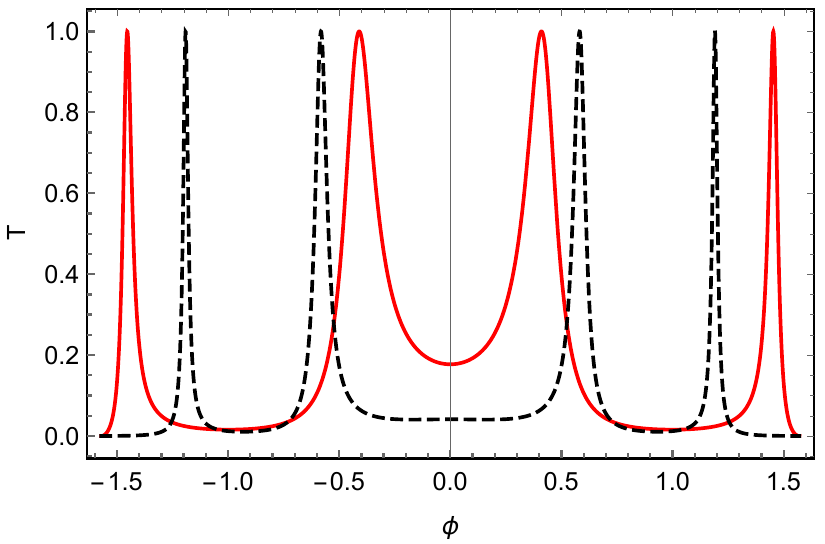}\label{fig4aa}}
	\\
	\subfloat[$\alpha=1$, n–p–n junction]{\includegraphics[scale=0.55]{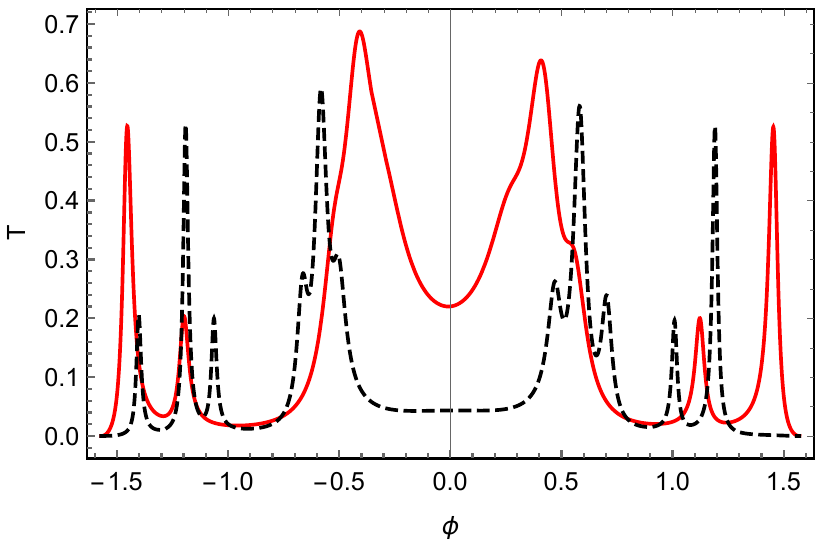}\label{fig4bb}}
	\\
	\subfloat[$\alpha=3$, n–p–n junction]{\includegraphics[scale=0.55]{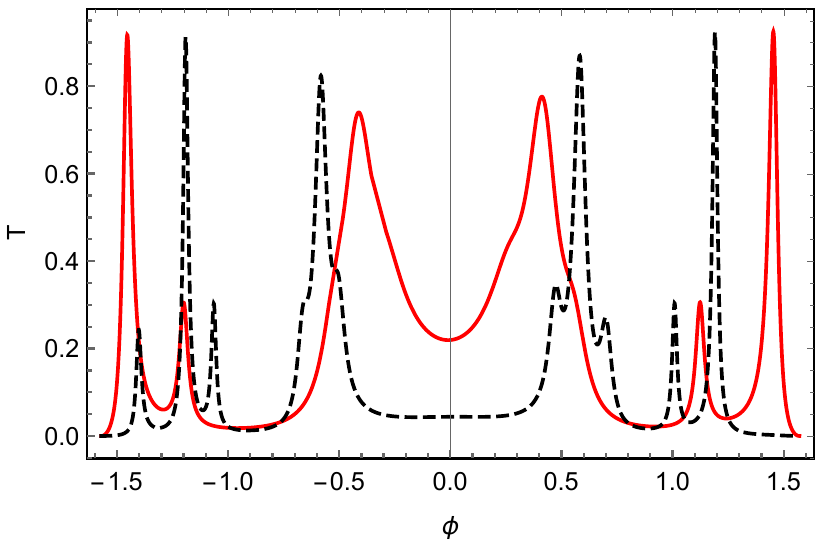}\label{fig4cc}}
	\caption{Total transmission spin-up  ($s_z=1$: red solid line) and spin-down ($s_z=-1$: black dashed line)
	 as a function of the incident angle $\phi$ in  $n$–$p$–$n$ junction for
	 three laser field parameters ($ \alpha=0, 1, 3$), 
	  $E=1.2$ eV, $V_0=2.6$ eV,  $\omega=12 \times 10^{12}$ Hz, and  $D=10$ nm.}
	\label{fig44} 
\end{figure}

Fig.~\ref{fig44} presents the total transmission  spin-up  (red solid line) and spin-down  (black dashed line) as a function of the incident angle $\phi$ {in the $n-p-n$ jonction for $k$ value}. In the absence of laser irradiation ($\alpha = 0$), Fig.~\ref{fig4aa} shows that the transmission probability is nearly zero for most angles, except for a few specific angles where the barrier becomes partially transparent. At normal incidence, spin-down fermions cannot pass through the barrier. This is known as the anti-Klein effect. It is the opposite of what happens in pristine graphene, where transmission is perfect \cite{klien1}. 
For spin-up fermions, transparency is not complete at normal incidence, and the transmission peaks observed at oblique angles are significantly shifted due to spin-orbit coupling, as demonstrated in the study \cite{CHENG2015}. For moderate laser intensity ($\alpha = 1$), Fig.~\ref{fig4bb} shows that overall transmission decreases. At the same time, new incident angles appear where transmission is no longer zero. This suggests that the laser reduces the range of angles for which the barrier is completely opaque. At low incident angles,  transmission spin-up remains stronger than  transmission spin-down. 
Under stronger laser irradiation ($\alpha = 3$), shown in Fig.~\ref{fig4cc}, the transmission spin-down increases at low angles and also transmission spin-up  rises. Laser intensity lets us control transmission based on the angle of incidence. This way, electron flow can be guided by changing both the light strength and the angle.

\begin{figure}[ht]
	\centering
\subfloat[$\alpha=0$, (n–n–n, p-p-p) junctions]{\includegraphics[scale=0.55]{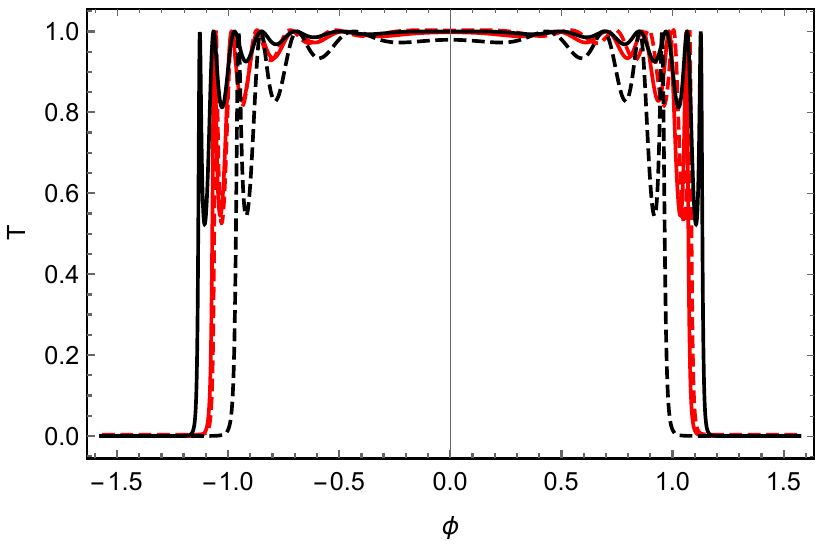}\label{fig45a}}
\\
\subfloat[$\alpha=1$, (n–n–n, p-p-p) junctions]{\includegraphics[scale=0.55]{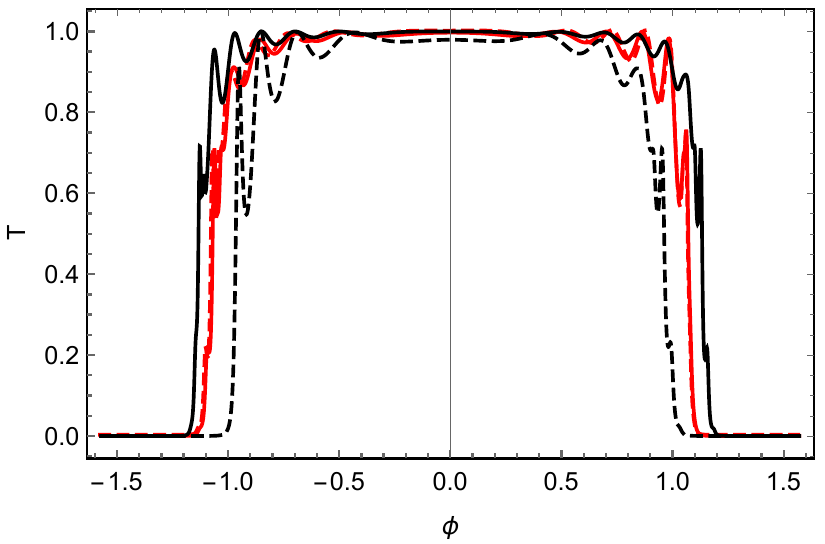}\label{fig45b}}
\\
\subfloat[$\alpha=3$, (n–n–n, p-p-p) junctions]{\includegraphics[scale=0.55]{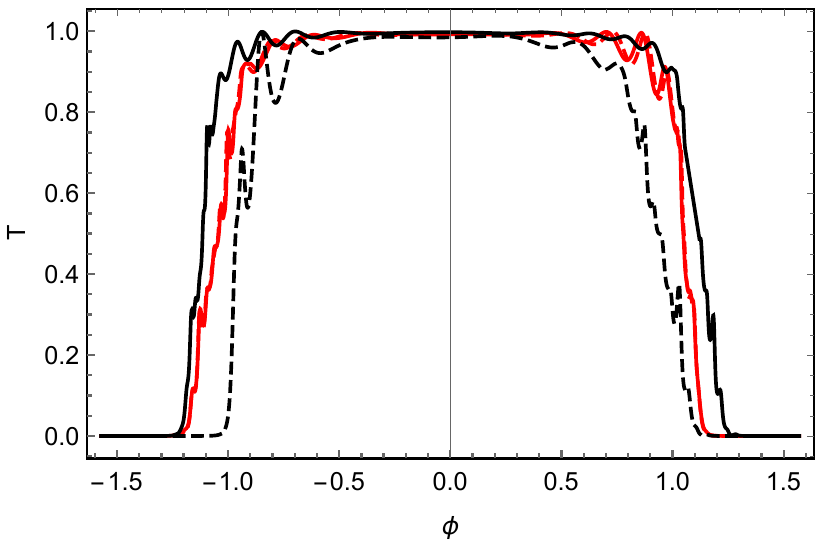}\label{fig45c}}
	\caption{Total transmission  as a function of the incident angle $\phi$ for n–n–n junction ($E = 1.2$ eV, $V_0 = 0.05$ eV) (red), p–p–p junction ($E = -1.2$ eV, $V_0 = -0.05$ eV) (black), three laser parameters ($\alpha = 0,1,3$),
		 $D = 10$ nm. The solid lines (dashed lines) correspond to spin-up (spin-down).}
\label{fig45} 
\end{figure}

Fig.~\ref{fig45} illustrates the transmission probability as a function of the incidence angle for the $p\text{--}p\text{--}p$ (black) and $n\text{--}n\text{--}n$ (red) junctions in the $K$ valley, distinguishing between spin-up electrons (solid lines) and spin-down electrons (dashed lines). Fig.~\ref{fig45a} corresponds to the case without laser irradiation ($\alpha = 0$). In the $n\text{--}n\text{--}n$ junction, the transmission curves for spin-up (solid red) and spin-down (dashed red) electrons are nearly identical, which can be attributed to the weak spin–orbit coupling in the conduction band \cite{CHENG2015}. In contrast, the $p\text{--}p\text{--}p$ junction exhibits a noticeable shift between the two curves, reflecting the stronger spin–orbit coupling in the valence band \cite{CHENG2015}.
For both junction types, the barrier remains largely transparent over a wide range of incidence angles. Conversely, in the $n\text{--}p\text{--}n$ configuration (see Fig.~\ref{fig4}), transmission is strongly suppressed at normal incidence, a phenomenon known as the anti-Klein effect. Fig.~\ref{fig45b}, plotted for $\alpha = 1$, shows that the transmission oscillations are reduced, while nearly total transmission persists for both the $n\text{--}n\text{--}n$ and $p\text{--}p\text{--}p$ junctions.
Finally, Fig.~\ref{fig45c} corresponds to the case of an intense laser field ($\alpha = 3$). Here, transmission remains maximal at normal incidence for both spin states in the $n\text{--}n\text{--}n$ and $p\text{--}p\text{--}p$ junctions. However, in the $n\text{--}p\text{--}n$ junction, the transmission amplitude becomes highly sensitive to the parameter $\alpha$, whereas its influence remains limited for the other two configurations.

\begin{figure}[ht]
		\centering
	\subfloat[$\alpha=0.5$]{\includegraphics[scale=0.55]{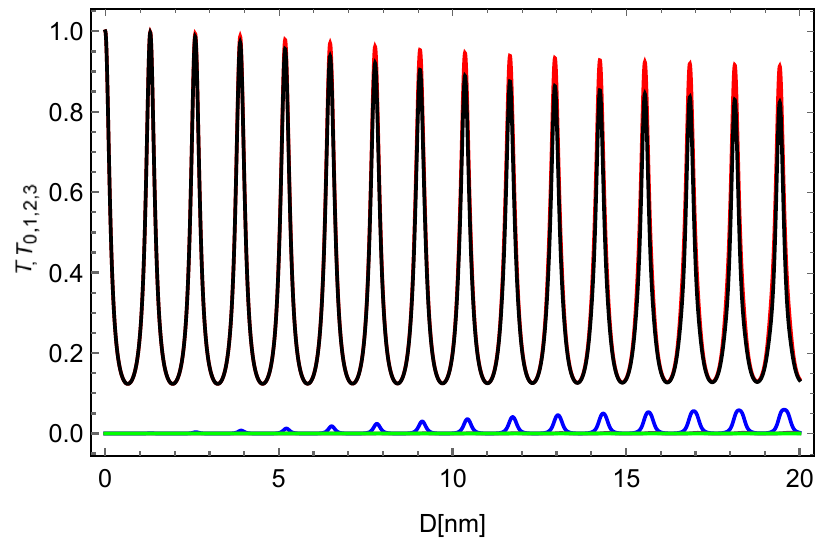}\label{fig2a}\includegraphics[width=0.7cm,height=4cm]{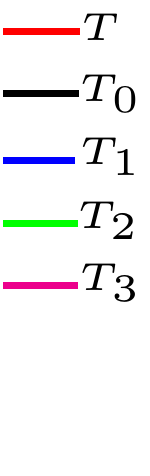}}\\
	\subfloat[$\alpha=1.5$]{\includegraphics[scale=0.55]{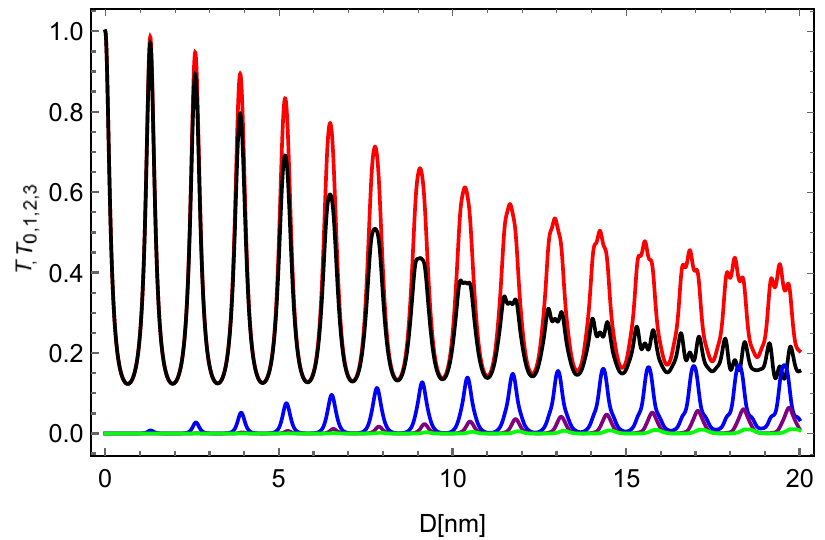}\label{fig2b}\includegraphics[width=0.7cm,height=4cm]{T0123.png}}\\
	\subfloat[$\alpha=3$]{\includegraphics[scale=0.55]{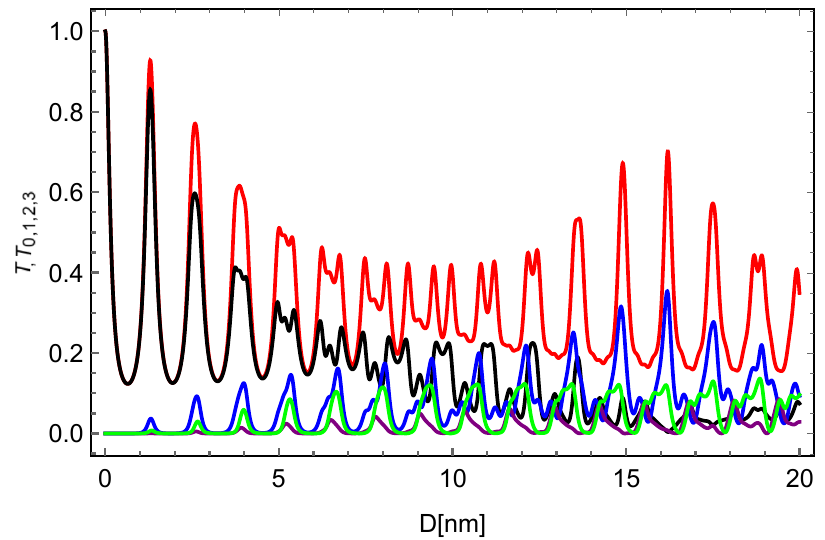}\label{fig2c}\includegraphics[width=0.7cm,height=4cm]{T0123.png}}
	\caption{Transmissions spin-up as a function of barrier width $D$ at normal incidence ($\phi=0$) for three laser parameters (\(\alpha=0.5, 1.5, 3\)),
		\(\omega = 12 \times 10^{12}\) Hz, \(V = 2.6\) eV, and \(E = 1.2\) eV.
		Here, \(T\) (red solid line), \(T_0\) (black solid line), \(T_1\) (blue solid line), \(T_2\) (green solid line), and \(T_3\) (magenta solid line).
	}\label{fig2} 
\end{figure}

Fig.~\ref{fig2} shows the transmissions $T_l$ $(l=0, \cdots, 3)$ and total transmission $T$ (red line) as a function of the barrier width $D$ for different values of $\alpha$. The transmission exhibits an oscillatory variation, with a dominance of the transmission $T_0$ without photon exchange. 
For $\alpha = 0.5$ in Fig.~\ref{fig2a}, we observe that $T_0$ is almost equal to the total transmission $T$. This means that most electrons pass through the barrier without exchanging photons. In this low-irradiation regime, photon-assisted processes are minimal, and the barrier remains transparent mainly without laser interaction. However, photon exchange increases with increasing barrier width. More distance allows electrons to interact with the laser field, increasing their ability to absorb or emit photons and gain energy.
In Fig.~\ref{fig2b}  for $\alpha=1.5$, we see that the effect of laser irradiation is very clear. Indeed, $T$ and $T_0$ decrease, while transmission with photon exchange becomes more probable. For more intense irradiation $\alpha=3$, Fig.~\ref{fig2c} shows that transmission with photon exchange becomes more dominant for a wide barrier. 
The enlargement of the barrier causes the fermions to interact for a longer time with the laser field, which enhances the probability that they will interact with the structure by the photon exchange. Through these exchanges, fermions can absorb more energy and thus increase the probability of tunnelling through the barrier. This mechanism explains the increase of the photon-assisted transmissions at high laser field intensity, which was also detected in graphene-based systems \cite{Elaitouni2023A,doublelaser}.

\begin{figure}[ht]
	\centering
	\subfloat[$\alpha =0.5$]{\includegraphics[scale=0.55]{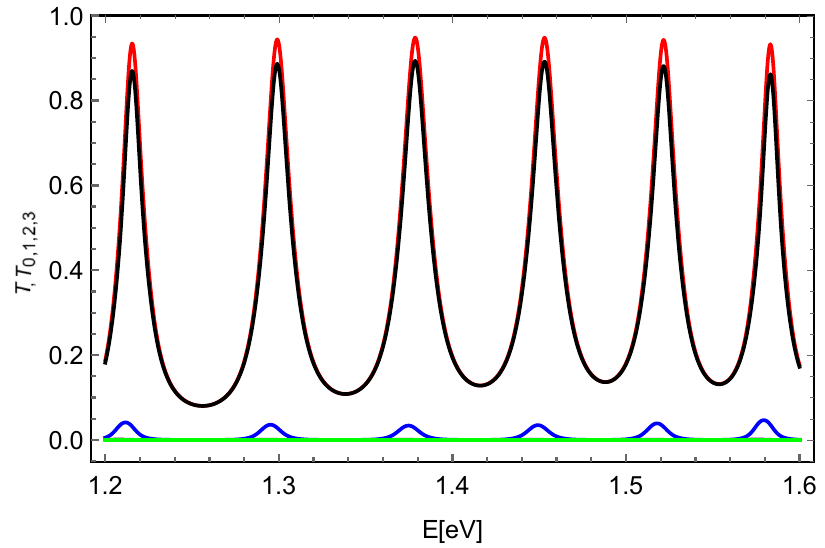}\label{fig3a}\includegraphics[width=0.7cm,height=4cm]{T0123.png}}\\
	\subfloat[$\alpha =1.5$]{\includegraphics[scale=0.55]{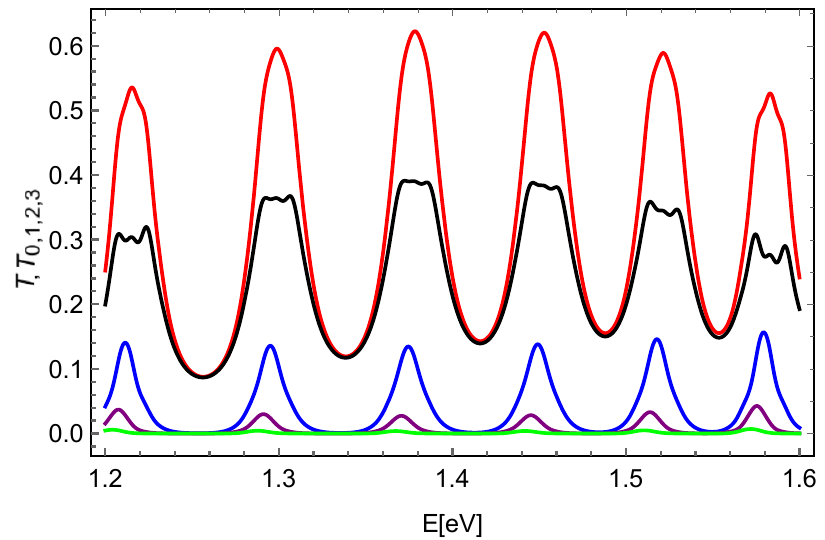}\label{fig3b}\includegraphics[width=0.7cm,height=4cm]{T0123.png}}\\
	\subfloat[$\alpha =2$]{\includegraphics[scale=0.55]{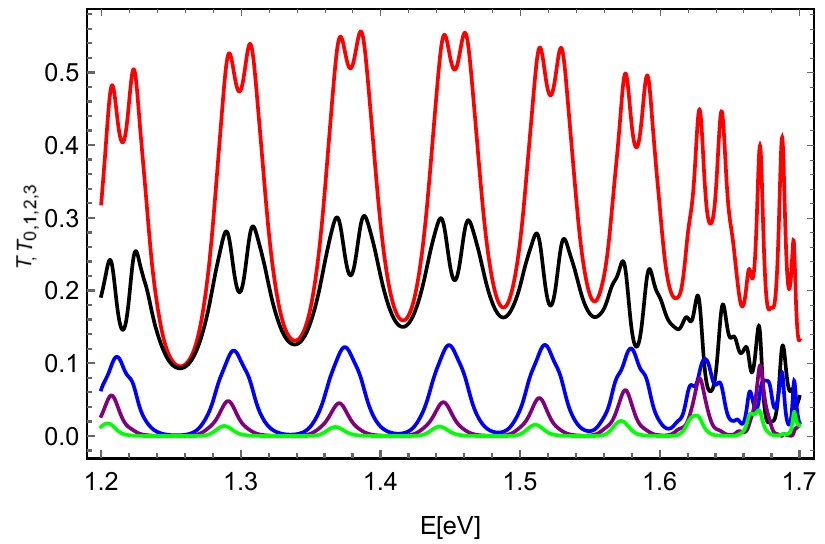}\label{fig3c}\includegraphics[width=0.7cm,height=4cm]{T0123.png}}
	\caption{Transmissions spin-up as a function of incident energy $E$ at normal incidence ($\phi=0$) for three laser parameters (\(\alpha=0.5, 1.5, 2\)),
		 \(D=10\) nm, \(V_0=2.6\) eV, and \(\omega=12.5\times10^{12}\) Hz.
		 Here, \(T\) (red solid line), \(T_0\) (black solid line), \(T_1\) (blue solid line), \(T_2\) (green solid line), and \(T_3\) (magenta solid line).}\label{fig3} 
\end{figure}

Fig.~\ref{fig3} shows transmission spin-up for the central band and the first three sidebands versus the  incident energy $E$. In Fig.~\ref{fig3a}, with a low laser strength ($\alpha = 0.5$), the transmission through the central band $T_0$ dominates. It closely matches the total transmission, which displays a nearly periodic pattern, reaching peaks at specific energies. As the laser strength increases to $\alpha = 1.5$ (Fig.~\ref{fig3b}), a noticeable change occurs. The total transmission $T$ decreases significantly and stays below $60\%$. At the same time, transmissions involving photon absorption or emission begin to grow, while $T_0$ drops. This trend becomes more pronounced at $\alpha = 2$, as shown in Fig.~\ref{fig3c}. Here, the effect of laser irradiation on electron transport becomes very strong. The direct transmission without photon exchange is further suppressed, while photon-assisted channels become dominant. These results highlight the ability of a laser field to reshape the fermionic energy spectrum. It promotes inelastic processes and modifies how electrons scatter. In short, higher laser intensity encourages photon-related interactions, which reduce the total transmission and introduce more complex transport behavior in the system.

\section{Conclusion}
We conducted a detailed analysis of the tunneling phenomenon in molybdenum disulfide MoS$_2$, focusing on its behavior when subjected to a static potential barrier in a region of width $D$. This region is irradiated by a monochromatic laser field with linear polarization. The introduction of this barrier divides the system into three distinct regions, creating a scenario where quantum transmission must be thoroughly examined. Given the temporal periodicity of the Hamiltonian induced by the laser field, we employ Floquet theory to solve for the wave functions within each region. The application of the wave function continuity condition at the barrier interfaces results in four fundamental equations, each admitting an infinite number of possible transmission modes. Initially, we derive analytical expressions for the transmission coefficients corresponding to the central band and the first lateral band, providing a fundamental understanding of how tunneling occurs under these conditions. For higher-order transmission modes, we implement the matrix formalism to obtain numerical solutions. 
The transmission properties of spin-up (spin-down) electrons in the $K$ valley are equivalent to those of spin-down (spin-up) electrons in the $K^\prime$ valley. This equivalence arises from the invariance of the Hamiltonian between the two valleys, expressed by the relation $H(K, \uparrow) = H(K', \downarrow)$.


Our numerical analysis demonstrates that laser illumination is a dominant factor in controlling the transmission properties of MoS$_2$. Specifically, it plays a critically important role in suppressing Klein tunneling—a relativistic quantum process in which particles exhibit perfect transmission through potential barriers due to their chirality. In addition, laser illumination is also important in controlling spin-selective transport through the barrier. The coupling of the fermions with the incident laser field alters the energy levels of the fermions, and this results in them being transmitted preferentially based on spin orientation. Spin-up fermions are better interacting with the field of irradiation and therefore have higher chances of penetrating the barrier compared to spin-down fermions. The probabilities of transmission vary based on various parameters like laser irradiation intensity, width of the barrier, and energy of incident fermions. At low irradiation intensities, transmission is controlled by the middle band with minimal spin selectivity. However, as the laser intensity increases, photon exchange gains more control and amplifies the distinction in transport behavior between spin-up and spin-down fermions. This phenomenon has profound consequences for spintronics and quantum transport applications, insofar as it means that externally generated spin-polarized currents and electronic conductivity electronically tunable can be achieved in two-dimensional semiconductors like molybdenum disulfide using laser fields. Knowledge of these underlying principles provides the pathway toward the realization of optically tunable spintronic devices as well as greater spin-dependent quantum transport control.

\end{document}